\documentclass[aps,prl,showpacs,twocolumn,superscriptaddress, floatfix, tightenlines, amsmath, amssymb,longbibliography]{revtex4-1}
\usepackage[pdftex]{graphicx} 

\usepackage{natbib}
\usepackage{booktabs}
\usepackage{color}
\usepackage{url}
\usepackage{float}
\usepackage{amsmath}
\usepackage{soul,xcolor}
\usepackage{hyperref}
\usepackage{nameref}
\usepackage{lineno}
\newcommand{\tas}{$1T$-TaS$_2$} 
\newcommand{\ticdw}{$T_{\mathrm{I-NC}}$}
\begin{document}

\title{Dynamic phase transition into a mixed-CDW state in 1T-TaS$_2$ via a thermal quench}

\author{Alberto de la Torre}
\email[Corresponding author:]{a.delatorreduran@northeastern.edu}
\affiliation{Department of Physics, Northeastern University, Boston, Massachusetts, 02115, USA}
\affiliation{Quantum Materials and Sensing Institute, Northeastern University, Burlington, Massachusetts, 01803, USA}
\affiliation{Department of Physics, Brown University, Providence, Rhode Island 02912, United States}
\author{Qiaochu Wang}
\affiliation{Department of Physics, Brown University, Providence, Rhode Island 02912, United States}
\author{Yasamin Masoumi}
\affiliation{Department of Physics, Northeastern University, Boston, Massachusetts, 02115, USA}
\author{Benjamin Campbell}
\affiliation{Department of Physics and Astronomy, University of New Hampshire, Durham, NH 03824.}
\author{Jake V. Riffle}
\affiliation{Department of Physics and Astronomy, University of New Hampshire, Durham, NH 03824.}
\author{Dushyanthini Balasundaram}
\affiliation{Department of Physics and Astronomy, George Mason University, Fairfax, VA, 22030 USA}
\affiliation{Quantum Science and Engineering Center, George Mason University, Fairfax, VA, 22030 USA}
\author{Pattrick M.Vora}
\affiliation{Department of Physics and Astronomy, George Mason University, Fairfax, VA, 22030 USA}
\affiliation{Quantum Science and Engineering Center, George Mason University, Fairfax, VA, 22030 USA}
\author{Jacob P.C. Ruff}
\affiliation{Cornell High Energy Synchrotron Source, Cornell University, Ithaca, NY, 14853, USA}
\author{Gregory A. Fiete}
\affiliation{Department of Physics, Northeastern University, Boston, Massachusetts, 02115, USA}
\affiliation{Quantum Materials and Sensing Institute, Northeastern University, Burlington, Massachusetts, 01803, USA}
\affiliation{Department of Physics, Massachusetts Institute of Technology, Cambridge, MA 02139, USA}
\author{Shawna M. Hollen}%
\affiliation{Department of Physics and Astronomy, University of New Hampshire, Durham, NH 03824.}
\author{Kemp W. Plumb}
\email[Corresponding author:]{kemp_plumb@brown.edu}	
\affiliation{Department of Physics, Brown University, Providence, Rhode Island 02912, United States}

\date{\today}

\maketitle

\textbf{Ultrafast light-matter interaction has emerged as a new mechanism to exert control over the macroscopic properties of quantum materials toward novel functionality. To date, technological applications of non-thermal phases are limited by their ultrashort lifetimes and low-ordering temperatures. The hidden metallic charge density wave (H-CDW) in \tas{} is among the most studied photoinduced metastable phases because of its technological promise. However, despite active study, the nature of the photoinduced H-CDW remains the subject of debate and potential applications have been limited because it has so far only been stabilized at cryogenic temperatures. Here, we stabilize the H-CDW phase at thermal equilibrium by accessing a mixed CDW order regime via thermal quenching. Using x-ray high dynamic range reciprocal space mapping (HDRM) and scanning tunneling spectroscopy (STS), we reveal the coexistence of commensurate (C) CDW and H-CDW domains up to 210 K. Our findings show that each order parameter breaks basal plane mirror symmetry with different chiral orientations and induces out-of-plane unit cell tripling in the H-CDW phase. Despite metallic domain walls and a finite density of states at zero bias observed via STS, bulk resistance remains insulating due to CDW stacking disorder. By comparing our data to Landau-Ginzburg theory calculations of the free energy functional, our study establishes the H-CDW as a thermally stable phase and introduces a new mechanism for switchable metallic behavior in thin flakes of \tas{} and similar materials with competing phases.}


\section{Introduction} 

The correlated van der Waals material \tas{} displays a series of first-order phase transitions associated with the onset of CDW orders with different degrees of commensurability \cite{THOMPSON1971,Wilson1975,fazekas1979electrical}. Lowering from high temperatures, at $T = 550$~K a mirror-symmetry preserving incommensurate (I) CDW onsets from a metallic state \cite{Sipos2008}. At \ticdw{}$ \approx 350 $~K, \tas{} enters into a nearly commensurate (NC) CDW phase. The atomic rearrangement associated with the incommensurate NC-CDW (in-plane ordering wavevector $q_{||} \!=\! 0.284$ r.l.u.) leads to disconnected regions displaying a  $\sqrt{13} \times \sqrt{13}$ spatial reconstruction in which twelve Ta atoms move towards a central one forming a star-of-David (SOD) pattern \cite{PhysRevB.56.13757}. The NC-CDW breaks the mirror symmetries of the lattice basal plane and spontaneously selects one of two chiral domains rotated from the $a$-axis by $\phi \approx 12^{\circ}$. Then, at $T_\mathrm{NC-C} \approx 175$~K in cooling and $T_\mathrm{NC-C} \approx 225$~K in warming, a 3D CDW order emerges from the NC-CDW phase. This low-temperature phase is commensurate (C) with the atomic lattice in-plane ($\phi \!=\! 13.9 ^{\circ}$ and $q_{||} \!=\! 0.277$ r.l.u.) with the in-plane C-CDW atomic reconstruction inheriting the chiral structure from the NC-CDW. On the other hand, in the out-of-plane direction, the C-CDW displays a large stacking disorder between dimerized CDW planes \cite{THOMPSON1971,Wilson1975}.

Experimental probes spanning a large range of length scales find that in the C-CDW phase of \tas{} samples lack the chiral twin domains expected across the $T_\mathrm{NC-C}$ first-order phase transition \cite{Zong_UED,Fichera_SHG_mirror_symmetry_2020, Liuyan_TaS2, Chiral_electronic_structure,Campbell}. The stabilization of a single chiral domain reflects the kinetics of the cooling across the only first-order phase transition in \tas{} that breaks in-plane mirror symmetries, \ticdw{} \cite{Sept_inhomogenous, Binder_1987}. During synthesis, the metastable $1T$ phase is stabilized by quenching from $T \approx 1000$~K to $T \approx 273$~K across the polytype transition, $T_{\rm{1T-2H}} \!=\! 600$~K  \cite{THOMPSON1971}. During this initial quench, the cooling rate across \ticdw{} is effectively slow when compared to the intrinsic timescales of the electronic ($\approx$ fs) and lattice ($\approx$ ps) degrees of freedom and enables the nucleation of large chiral CDW domains with lateral sizes comparable to the correlation length of the order parameter parameter \cite{Oxtoboy_1998}. The chiral mono-domain structure of the C-CDW in \tas{} is stable to cooling through $T_{\mathrm{NC-C}}$ \cite{Zong_UED,Fichera_SHG_mirror_symmetry_2020,Liuyan_TaS2,Chiral_electronic_structure,Campbell} with suggestions that strain fields and defects can pin a single chirality \cite{Zong_UED, PhysRevMaterials_2023_salzmann}. Still, chirality reversal of single domain samples has been demonstrated by repeated cleaving of a bulk sample \cite{Fichera_SHG_mirror_symmetry_2020} or by cycling through \ticdw{} \cite{Zhao2023}. 

On the other hand, multi-chiral domain samples can be engineered in thin exfoliated flakes by heating and quenching above $T_{\rm{1T-2H}}$ \cite{Sung2022,Husremović2023} or upon irradiation with 800 nm laser pulses with fluence $F > 7$~mJ/cm$^2$ \cite{Zong_UED}. Once created, chiral domain walls are long-lived at low temperatures, but a dominant chiral domain of dimensions comparable to the sample size can be restored by heating above \ticdw{} and adiabatically cooling through the mirror symmetry-breaking IC-NC transition \cite{Sung2022,Husremović2023,Zong_UED}. Thus, \ticdw{} emerges as the relevant energy scale for chiral CDW domain nucleation in \tas{} as the temperature at which the initial mirror-symmetry breaking transition in the CDW phase diagram \cite{THOMPSON1971} occurs and the largest specific heat divergence is observed \cite{Manas-Valero2021}. Moreover, x-ray pair distribution function measurements have revealed that the anisotropic Ta Debye-Waller factor, associated with CDW formations,  remains constants across the NC-C CDW transition but dramatically changes across \ticdw{} \cite{bozin2023crystallization}. 

\begin{figure*}[!ht]
    \includegraphics[width=0.75\textwidth]{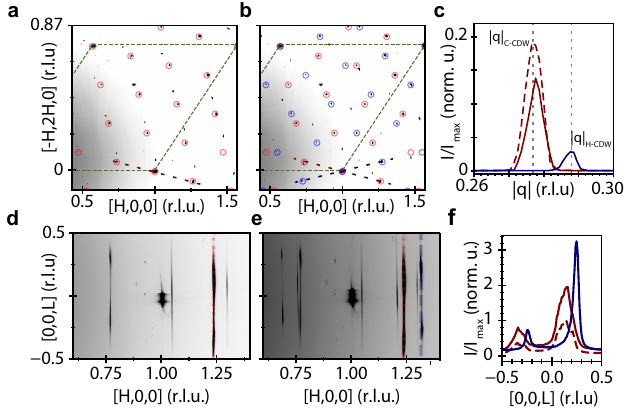}
	\caption{\textbf{Thermal quench into the mixed-CDW phase.} \textbf{a}. Reciprocal space maps parallel to the [HK0] plane ($\delta L \!=\! .5$, L \!=\! 0) at $T \!=\! 140$~K for an as-grown sample. \textbf{b}. Reciprocal space maps parallel to the [HK0] plane ($\delta L \!=\! .5$, L \!=\! 0) at $T \!=\! 80$~K after a quench. Red and blue markers highlight the C-CDW and H-CDW-like satellite peaks, respectively. Green dashed lines highlight the $a^{*}$ and $b^{*}$ directions. \textbf{c}  Average intensity dependence as a function of $|q|$ across a C-CDW satellite peak in the as-grown sample (dashed red line) and the C-CDW (red line) and H-CDW-like (blue line) after a quench. (Red and blue dashed lines in \textbf{b}). Vertical dashed lines correspond to the reported $|q_{||}|$ values for the C-CDW \cite{THOMPSON1971} and H-CDW \cite{Stahl_NatComm_XRD} satellite peaks.  \textbf{d} Reciprocal space map parallel to the [H0L] plane for the as-grown sample. \textbf{e} Reciprocal space map parallel to the [H0L] plane for the quench sample. Vertical red and blue dashed lines indicate C-CDW and H-CDW-like satellite peaks, respectively. \textbf{f} Integrated intensity along [h0L] for the as-grown sample (red dashed) and after a quenched (red, C-CDW peak and blue, H-CDW peak).}
	\label{fig:f1}
\end{figure*}

An additional low-temperature metastable metallic CDW phase, the so-called hidden H-CDW phase, has been accessed in \tas{} from the insulating C-CDW state via a single femtosecond laser pulse of fluence larger than a critical value ($F
_c\!>\! 1$~mJ/cm$^2$) \cite{Stojchevska_Science_2014,Stahl_NatComm_XRD,Vaskivkyi_2015} or by exciting \tas{} with a ns voltage pulse \cite{hollander2015electrically,ma2016metallic,cho2016nanoscale,Venturini_2022,Tsen_pnas}. The H-CDW phase locally resembles the SOD arrangement but breaks the large C-CDW domains into many nanoscale domains and has a metallic density of states \cite{Gerasimenko2019,ma2016metallic,Retting_2023}. These domains maintain the chirality of the C-CDW \cite{Stahl_NatComm_XRD} and are related by a relative translation of the SOD pattern by a finite number of atomic lattice vectors \cite{ma2016metallic,Gerasimenko2019,ravnik2021time}. Slight differences between the displacement of the Ta atoms between the H-CDW and the C-CDW distinguish the two. The H-CDW phases result in CDW Bragg peaks rotated by $\phi \!=\! 11.9^{\circ}$ from the $a^*$-axis with a larger basal plane ordering wavevector, $q_{||} \!=\! 0.284$ r.l.u.. The CDW plane stacking also rearranges towards a $ l \!=\! 1/3$ periodicity along the c-axis \cite{Stahl_NatComm_XRD}. We note that a similar metallic mosaic state can also be induced by cooling thin exfoliated samples ($d < 24$~nm) across \ticdw{} with a cooling rate faster than $0.2$~K/s \cite{Yoshida2015} or by intrinsic strain and defect effects \cite{Mutka1981, Zhang2022,Salzmann2023}. 

Despite much study, the origin of the H-CDW state remains controversial, with photodoping \cite{KS_photodoping,Gerasimenko2019}, incoherent thermal effects \cite{ma2016metallic,Gao_Timescales} or electron-phonon relaxation bottlenecks \cite{Retting_2023,mante2022photoinduced} all possible explanations.  Nonetheless, the H-CDW has attracted much interest, not only as an example of the non-thermal control of quantum materials \cite{RevModPhys.93.041002} but also as a platform for new functional devices \cite{tas2_devices, Mraz2022, devidas2024spontaneous}. However, the H-CDW has only been stabilized for ultrashort periods ($\sim100\,\, \mu$s)  at temperatures above $T \!=\! 50$~K \cite{Stojchevska_Science_2014, ravnik2018real, Retting_2023} preventing its deployment in efficient and fast memory devices towards novel technological applications \cite{basov2017towards}.

Here, we demonstrate that the H-CDW phase exists in thermal equilibrium in \tas{} and extend its thermal stability by a factor of three by accessing a new long-lived mixed-CDW state via a reversible thermal quench across \ticdw{}. We use x-ray high dynamic range reciprocal space mapping (HDRM) to unequivocally demonstrate the co-existence of the C-CDW and H-CDW states in thermally quenched samples. Each CDW order parameter is linked to a mirror-symmetry breaking chiral domain. The mixed-CDW state emerges from a multi-domain NC-CDW state below $T^{CD} = 180$~K in cooling and $T^{WU} = 210$~K in warming. The mixed-CDW low temperature state can be recovered as long as the system is not heated above $T_{I-NC}$. Scanning tunneling spectroscopy (STS) shows that the mixed-CDW is characterized by a semi-metallic in-plane density of states. At the same time, bulk resistivity remains insulating due to out-of-plane CDW stacking disorder in mixed CDW bulk samples. Our results and calculations of the free energy landscape within the Ginzburg-Landau theory demonstrate a new thermally accessible state in the phase diagram of \tas{}, enabling a new mechanism to engineer a semimetallic density of states at temperatures well above liquid nitrogen temperatures.
\begin{figure*}[!ht]
    \includegraphics[width=.98\textwidth]{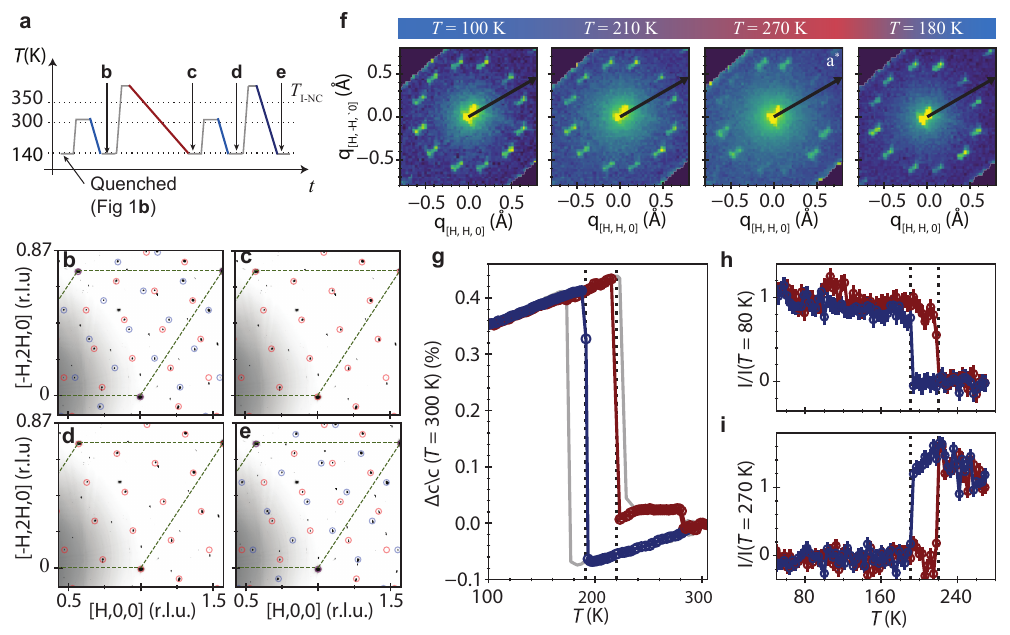}
	\caption{\textbf{Temperature dependence of the mixed-CDW.}\textbf{a} Schematic of the subsequent in-situ thermal cycles undergone by a pre-quenched sample. \textbf{b-e} HRDM reciprocal space maps parallel to [HK0] plane ($\delta L \!=\! .5$, L \!=\! 0) at $T \!=\! 140$~K after a thermal cycle as indicated in \textbf{a}. Red and blue markers highlight the C-CDW and H-CDW-like satellite peaks. Green dashed lines highlight the $a^{*}$ and $b^{*}$ directions. \textbf{f} Ag $K_{\alpha}$ reciprocal space maps parallel to the [HK0] ($\delta L \!=\! .5$, L \!=\! 10) as a function of temperature across the C-NC transition. The black arrow encodes the $a^{*}$ direction. \textbf{g} Temperature dependence of the relative change of the $c$-axis lattice parameter during cooling (blue) and heating (red) for a quenched sample (circular markers). Grey solid lines showcase the same magnitude for an as-grown sample. The quenched crystal shows a narrow hysteresis window for the C-NC transition compared to an as-grown sample. \textbf{h} Temperature dependence of a H-CDW-like satellite peak ($L = 9.66$,$\delta L \!=\! .1$) during cooling (blue) and warming (red) The thermal hysteresis follows that of the $c$-axis. \textbf{i} Temperature dependence of a H-CDW-like-precursor satellite peak ($L = 9.33$,$\delta L \!=\! .1$) in the NC phase during cooling (blue) and warming (red). Error bars in \textbf{h} and \textbf{i} represent one standard deviation.}
	\label{fig:f2}
\end{figure*}

\section{Results}

\textbf{Accessing a novel mixed-CDW state by quenching through the mirror symmetry breaking transition.} In Figure \ref{fig:f1} \textbf{a}, we show a characteristic [HK0] plane reciprocal space map for the C-CDW phase of \tas{}, collected after a standard cooling cycle. Red circular markers highlight the CDW reflections at  $q^{C}_1 \!=\! (\sigma^{C}_1,\sigma^{C}_2,l)$ and $q^{TH}_2 \!=\! (-\sigma^{C}_2,\sigma^{C}_1+\sigma^{C}_2+,l)$ with $\sigma^{C}_1 \!=\! 3/13$ r.l.u and $\sigma^{C}_2 \!=\! 1/13$ r.l.u  \cite{THOMPSON1971,Wilson1975}, corresponding to a single C-CDW chiral domain. Additional non-indexable reflections in the reciprocal space map are attributed to a second small crystallite in our sample. Identical reciprocal space maps were obtained across many cooling cycles and samples.  We found that the C-CDW domain structure in a given sample depends on the cooling rate through \ticdw{} and fast thermal quench across \ticdw{} stabilizes a multi-chiral domain CDW. Fig.~\ref{fig:f1} \textbf{b} shows the [HK0] reciprocal space map after an in-situ quench in our bulk sample across \ticdw{} down to $T \!=\! 140$~K at a rate of $0.2$~K/ms near \ticdw{} (See Methods for a description of the in-situ quench method and calculation of quenching rates using heat diffusion calculations). We found that thermal quenching consistently resulted in a new set of superlattice reflections of opposite chirality to the original set of satellite peaks ($\phi \!=\! - 12.3^{\circ}$) (Fig. \ref{fig:f1} \textbf{b}, highlighted by blue markers). The superior reciprocal space resolution of synchrotron single crystal HDRM enables us to resolve differences in the wavevector of the two sets of CDW Bragg peaks with different chirality. The in-plane wavevector magnitude $|q_{||}| \!=\! 0.284(3)~\AA^{-1}$ of the quenched-induced satellite peaks is larger than that of the C-CDW ($|q_{||}| \!=\! 0.277(3)~\AA^{-1}$) and is the same as the light-stabilized H-CDW state \cite{Stahl_NatComm_XRD}. This is shown in Fig. \ref{fig:f1} \textbf{c}, where we display the integrated intensity across the C-CDW and H-CDW Bragg peaks (dashed lines in Fig. \ref{fig:f1} \textbf{a} and \textbf{b}) as a function of $|q_{||}|$. Strikingly, each order parameter preferentially selects a chirality orientation, with the C-CDW domains of the mixed-CDW phase maintaining that of the as-grown samples. 

We also find that the c-axis periodicity differs between the C-CDW and H-CDW-like domains. Figure \ref{fig:f1} \textbf{d} and \textbf{e} show [H0L] reciprocal space maps centered at H = 1 r.l.u. and L = 0 r.l.u. ($\delta k = 0.3$ r.l.u.) before and after the thermal quench, respectively. The as-grown sample displays a broad two-peak structure, with the center of mass at integer values of L. These short-range out-of-plane correlations are characteristic of the dimerization of CDW layers in the low-temperature C-CDW phase \cite{Tanaka_JPL_1984}. On the other hand, while retaining a dumbbell-like shape, the H-CDW-like satellite peaks shift towards $l \!=\! \pm 1/3$ indicating a collapse of the interlayer dimerization towards a triple layer CDW stacking as was observed for the light-induced H-CDW \cite{Stahl_NatComm_XRD}. The H-CDW-like satellite peaks are sharper than the C-CDW ones, reflecting a longer correlation length for the quench-induced domains $\xi \!=\!6(6)$ nm (Fig. \ref{fig:f1} \textbf{f}). Similar sharpening of the out of plane CDW peaks was also observed in light-induced H-CDW \tas{} \cite{Stahl_NatComm_XRD}. Altogether, our HDRM data in quenched samples demonstrates the coexistence of the C-CDW with a second CDW phase with the same order parameter as the H-CDW phase at much higher temperatures ($T = 140$~K). 

\textbf{Critical temperature and metastability of the quenched mixed-CDW state.} We performed a set of in-situ thermal cycles to directly demonstrate that (i) \ticdw{} is the relevant temperature scale to access the mixed state and (ii) the transition is reversible. Consecutive thermal quenches were performed at different quench rates and starting temperatures, as illustrated in Figure \ref{fig:f2} \textbf{a} (see Methods and Extended data). We found that after quenching from above \ticdw{}, the mixed CDW state is thermally stable and insensitive to warming-cooling cycles if temperature kept less than \ticdw{} [Fig.~\ref{fig:f2} \textbf{b}]. Such thermal stability is in contrast with the light-induced H-CDW phase that was found to be suppressed above $T \!=\! 60$~K \cite{Stojchevska_Science_2014}. However, the C-CDW state can be recovered from the mixed state by warming up and slow-cooling across \ticdw{} (Fig. \ref{fig:f2} \textbf{c}). Subsequent thermal quenches from $T \!=\! 300$~K do not recover the mixed state (Fig. \ref{fig:f2} \textbf{d}), which can only be recovered upon heating and quenching from above  \ticdw{} (Fig. \ref{fig:f2} \textbf{e}). Thus, our data confirms the IC-NC CDW as the necessary energy scale to access the mixed-CDW phase containing both the C-CDW and H-CDW order parameters. 

\begin{figure}[!t]
 \includegraphics[width=.5\textwidth]{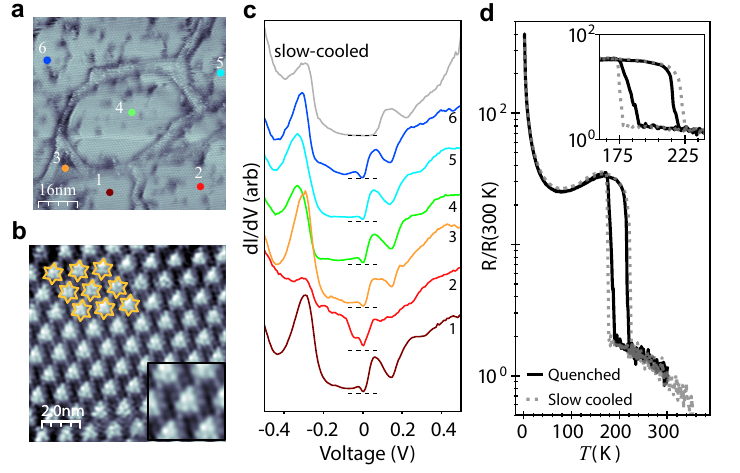}
	\caption{\textbf{Electronic structure of the mixed-CDW phase} \textbf{a -b} Representative topography of the mixed-CDW phase showing domains, atomic resolution and the $\sqrt{13} \times \sqrt{13}$ reconstruction. \textbf{c} $dI/dV$ curves at each of the 6 domains seen in \textbf{d}. $dI/dV$ data reflects a narrower gap in the electronic structure of \tas{} compared to the slow-cooled C-CDW phase (grey). Metallic behavior is observed in domain walls as indicated by point 3. \textbf{d} Resistivity, normalized to that at $T \!=\! 300$~K for a mixed-sample (solid line) and as-grown sample (dotted grey). The inset shows a zoom in view around the TH-CDW/C-CDW transition.}
	\label{fig:f3}
\end{figure}

The possibility of recovering the H-CDW phase as part of the stable mixed-CDW phase by keeping the samples below \ticdw{} enables us to ex-situ form the CDW mixed state (see Methods) and perform additional characterization of this state. We note that the mixed CDW state is stable over months, independent of the number of cleaves, and has been reproduced in quenched samples over multiple growth batches. Detailed temperature-dependent XRD measurements (See Methods) of ex-situ quenched samples display twelve satellite reflections characteristic of the mixed-CDW state at low temperatures (Fig.\ref{fig:f2} {f}). For temperatures above $T^{WU}_{NC-Mixed} \!=\! 210$~K during warming up, or below $T^{CD}_{NC-mixed} \!=\! 195$~K during cooling down, we observe a transition into the NC-CDW. However, the NC-CDW state in quenched samples is characterized by two sets of CDW Bragg peaks with opposite chiralities. This state is shown in Fig. \ref{fig:f2} \textbf{f} at $T = 270$~K for ex-situ quenched samples. Equivalent HDRM data at $T = 300$~K is shown in Extended Data \ref{fig:E2}. We note that the L dependence of the two NC-CDW satellite peaks is different from that of the H-CDW we observe at low temperature (see Extended data \ref{fig:E2}), ruling out a supercooled NC-CDW phase in our quenched samples \cite{Yoshida2015}. The existence of two chiral domains in the NC-CDW is consistent with previous ultrafast electron diffraction experiments that demonstrated the chirality of the low-temperature C-CDW domains was inherited from the symmetry-breaking at the IC-NC transition \cite{Dynamics_NC_IC} and that two NC-CDW domains were stabilized after laser excitation at room temperature \cite{Zong_UED}.

A detailed temperature dependence of the H-CDW-like and NC-CDW Bragg reflections are shown in Fig. \ref{fig:f2} \textbf{h} and \textbf{i}. Upon stabilizing the mixed state at low temperatures via a quench across $T_{I-NC}$, the H-CDW peaks directly emerge from NC-CDW peaks of corresponding chirality. This temperature dependence is distinct from the metastable H-CDW phase, which is suppressed above $T = 60$~K. \cite{Stahl_NatComm_XRD, Stojchevska_Science_2014, Gerasimenko2019}. We found that C-CDW and H-CDW-like Bragg peaks show the same temperature dependence (see Extended data) suggesting the coupled nature of the two order parameters in the mixed-CDW phase. The asymmetrically narrower hysteretic loop of the first-order transition, when compared to as-grown samples, $T^{CD}_{NC-C} \!=\! 180$~K and $T^{WU}_{NC-C} \!=\! 210$~K, in conjunction with the narrower CDW Bragg peaks along L and the temperature dependence of the [0010] structural Bragg peak shown in Fig. \ref{fig:f2} \textbf{c}, suggests longer range interlayer correlations and a more 3D nature of the H-CDW phase.  

Scanning tunneling microscopy (STM), measurements confirm the multi-domain nature of our quenched samples. Figure \ref{fig:f3} \textbf{a} shows a characteristic $80 \times 80$~nm$^2$ area in a quenched sample. Each domain shows the atomic reconstruction of the sulfur layer due to a low-temperature CDW, as shown in Fig. \ref{fig:f3} \textbf{c} with $V \!=\! 380$~mV, $I \!=\! 0.35$~nA, scale bar length $\!=\! 2$~nm). We note that the in-plane momentum resolution of our STM experiments prevents us from distinguishing between the C-CDW and the H-CDW-like domains. However, from the Fourier transform of STM images with atomic resolution, we extract a rotated satellite peak $\phi \!=\! 12.8 (5)^{\circ}$, that reflects the presence of domains of opposite chirality (related by a mirror operation along one of the high symmetry crystallographic directions) (See extended data). Our STM data resembles the in-plane inhomogeneity observed in optically and electrically driven \tas{}, where multiple domains of opposite phase have been reported \cite{Gerasimenko2019,ma2016metallic,Retting_2023}. We note that the smaller domain density prevented us from performing a comprehensive in-plane phase configuration analysis of the chiral domains, but we cannot exclude the presence of phase domains (related by a translation operation) \cite{Gerasimenko2019,ma2016metallic,Retting_2023}.

To summarize, intermediate-quenched bulk \tas{} samples host a mixed CDW state with two distinct order parameters, locked to opposite chiral phases, which emerge from a multi-chiral NC-CDW phase. Moreover, our temperature-dependent data reveals the narrowing of the hysteretic window across the NC-mixed-CDW phase transition. This indicates a stronger out-of-plane electronic correlation in the mixed-CDW phase of \tas{}.

\textbf{Electronic structure of the bulk quenched samples.} We now bring attention to the low-temperature electronic structure of the mixed-CDW phase. While the origin of the insulating nature of the C-CDW remains controversial \cite{Pettochi_Mott_vs_HG_2022,Wang2020_TaS2_arpes_T}, it is well established that the c-axis stacking of CDW domains strongly influences the electronic structure of \tas{} \cite{PhysRevB.98.195134,PhysRevLett.122.106404,butler2020mottness,ritschel2015orbital}. Moreover, recent STM and scanning tunneling spectroscopy (STS) studies of the stacking shift between in-plane domains in consecutive layers have identified interlayer coupling as the key ingredient that determines the band structure of the low-temperature CDW phase in \tas{} \cite{cho2017correlated,ma2016metallic,Zoology_DW,Gerasimenko2019,park2023stacking}. Thus, we use STS to study the low-temperature electronic state of the quenched samples. $dI/dV$ curves taken at $T \!=\! 10 $~K within six different domains, Fig. \ref{fig:f3} \textbf{a}, reveal a reduction of the $E_g\! = \! 300$~meV gap of the C-CDW \cite{Campbell} towards a state with a reduced but non-zero density of states at the Fermi level (Fig. \ref{fig:f3} \textbf{c}). While a transition to a semimetallic state is consistent with that observed for the H-CDW \cite{ma2016metallic,Gerasimenko2019}, the density of states maintains two pronounced peaks at around $V_B = -300$~meV and $V_B = 150$~meV, that have been associated with the lower and upper Hubbard bands \cite{PhysRevB.71.153101,PhysRevLett.73.2103,PhysRevB.92.085132}. Despite the finite in-plane density of states at $E_F$ and the formation of metallic domain walls (Fig. \ref{fig:f3} \textbf{b}, curve 3), electrical transport in the mixed-CDW sample resembles that of the as-grown crystals. 

\begin{figure*}[!t]
    \includegraphics[width=\textwidth]{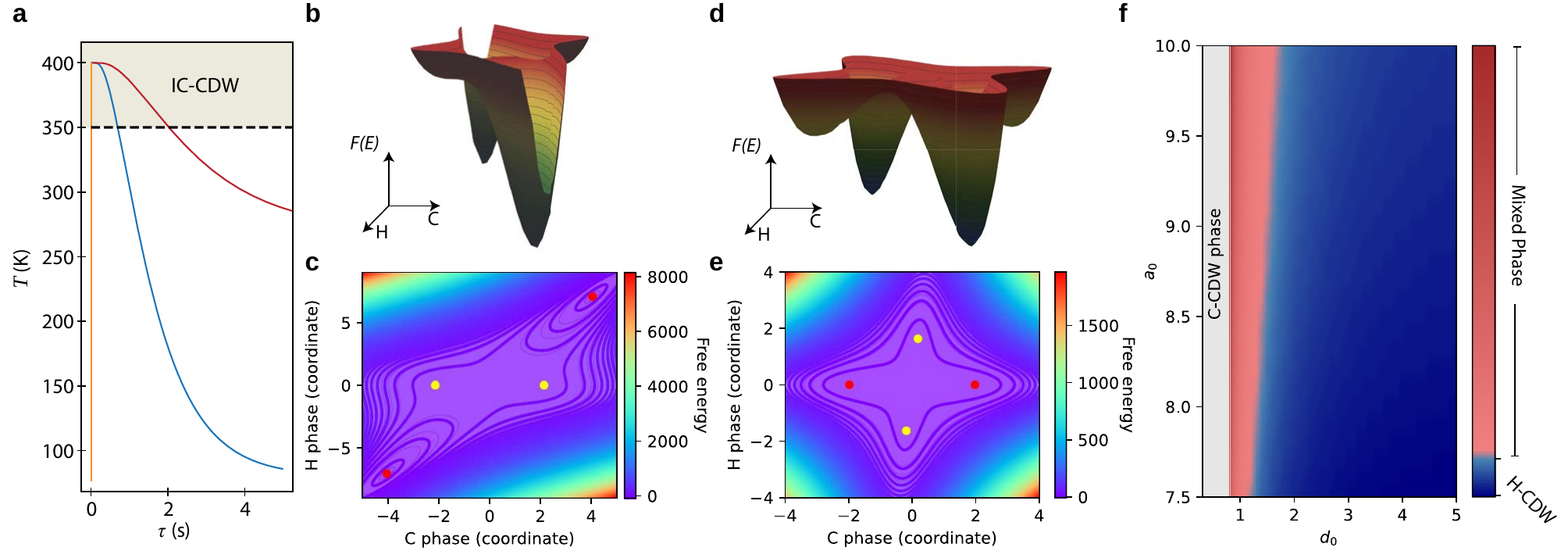}
	\caption{\textbf{Dynamic phase transition with two competing order parameters in \tas{}}. \textbf{a} Cooling rate estimates for an optically induced quench (orange) and solutions of the heat diffusion equation for the slow (red) and fast (blue) cooling schemes used in this work (See Methods and Extended data). \textbf{b} Meanfield-only free energy functional for the two competing CDW orders, with a competition factor $d_0 = 0.8$, corresponding to an adiabatic quench. \textbf{c} 2D projection of the energy functional in \textbf{b}. \textbf{d} Meanfield-only free energy functional for the two competing CDW orders, with a competition factor $d_0 = 6.1$, corresponding to laser quench.  Fast quenches correspond to larger values of $d_0$. \textbf{e} 2D projection of the energy functional in \textbf{d}. The yellow and red markers indicate local and global minima, respectively. \textbf{f} Computed phase diagram as a function of the competition factor, $d_0$, and mass term, $a_0$. The mixed CDW phase can only be stabilized in a narrow range of intermediate-rate quenches. Faster quenches (optical or electrical pulses) lead to the suppression of the C-CDW phase and the emergence of the H-CDW phase}.
	\label{fig:f4}
\end{figure*}

We observed that quenched bulk samples display insulating behavior at all temperatures with the same relative resistivity increase with respect to $R(300$~K) as the C-CDW phase \cite{THOMPSON1971} (Figure \ref{fig:f3} \textbf{d}). The mixed state resistivity exhibits a narrow thermal hysteresis that tracks that of the structural transition (Fig. \ref{fig:f2} \textbf{g}), reflecting the role of out-of-plane order in the resistivity of \tas{}.  The insulating nature of the mixed-CDW phase is distinct from the substantial $T<70$~K resistivity drop observed for the light-induced H-CDW phase in thin samples \cite{Stojchevska_Science_2014}. This difference likely originates from the insulating C-CDW domains that occupy a large volume fraction of the mixed-CDW state and stacking CDW disorder between the two CDW orders with different periodicity along L (Figure \ref{fig:f2} \textbf{g-h}). Additionally, we find that in-situ thermal cycling above $T_{IC} \!=\! 350$ ~K and slow cooling recovers the characteristic resistivity of bulk \tas{} with a hysteretical transition separated by $\Delta T \!=\! 50$~K (Figure \ref{fig:f3} \textbf{d}), consistent with our diffraction measurements in as-grown samples. Thus, our STS, resistivity, and XRD data support an interpretation in which bulk metallic transport is impeded by the stacking disorder along L of two order parameters with different long-range intralayer CDW order \cite{liu2024nonvolatile}.

\section{Discussion}


In summary, our results demonstrate that a mixed-CDW phase characterized by C-CDW and H-CDW domains of opposite chirality can be stabilized by a thermal quench through \ticdw{}. The required thermal quench is at an intermediate rate ($\approx 120$~K/s) based on heat diffusion calculations (see Methods), faster than the initial quench that \tas{} crystals experience during growth ($\leq 2$~K/s) but slower than the changes to the electronic temperature induced by a laser pulse ($\geq 1$ K/fs) \cite{Zong_UED,Stojchevska_Science_2014} (Figure \ref{fig:f4}) \textbf{a}. We note that the chiral splitting (relative rotation) of the two order parameters enables us to clearly resolve the mixed-CDW phase. If the H-CDW and C-CDW phases exhibited the same chirality, the small difference in their ordering wavevectors would lead to overlapping satellite peaks, complicating the observation of this novel state. Our STS data reveals that this mixed state hosts a semimetallic in-plane electronic density of states, while out-of-plane CDW stacking disorder prevents bulk metallic behavior. 

It is compelling to compare our observations with theories for dynamical phase transitions \cite{PhysRevX.10.021028,masoumi2024} and how domain formation influences the competition between coexisting order parameters. Thus, to describe the low temperature state after a quench, we consider a mean-field energy functional \cite{mcmillantheory} within the Landau-Ginzburg theory including, among others (see the Supplementary Information for more details of the calculations), a $d_0\bar{\phi}^2_C\bar{\phi}^2_H$ term in which $d_0$ parameterized the competition between the C-CDW and the H-CDW phase. The order parameter is the spatial charge modulation, $\alpha(\vec{r})=\Re(\phi_Ce^{i\vec{Q}_C\cdot\vec{r}}+\phi_He^{i\vec{Q}_H\cdot\vec{r}})$ in which the amplitude of the order parameter for the C-CDW, ${\phi}_C$, corresponds to the global minimum of the free energy landscape for slow cooling protocols, and $\phi_H$ is the amplitude of the H-CDW order parameter. We note that our model does not include the order parameter for the NC-CDW as the initial quench from \ticdw{} down to low temperature effectively skips over the intermediate CDW state. Morevoer, the H-CDW phase can only be stabilized from the C-CDW phase \cite{Stojchevska_Science_2014} indicating that the NC-CDW is not necessary to discuss the competition between these two states. Such free energy landscape allows a mixed inhomogeneous phase to be accessed at intermediate but adiabatic quenching rates in which a single temperature can be defined for all the different degrees of freedom in the material. This state is characterized by the coexistence of trapped regions of the two competing orders \cite{PhysRevB.93.125131} and enhanced lifetime due to the transiently created domain walls \cite{PhysRevX.10.021028}.

These behaviors are illustrated in Fig. \ref{fig:f4}\textbf{b-c} where the energy functional, $F(E)$, is calculated for $d_0 = 0.8$ with the same mass term, $a_0$, for both order parameters, $\frac{a_0}{2}\int dr^2\alpha(\vec{r})^2=\frac{a_0}{2}(\bar{\phi}_C^2+\bar{\phi}_H^2)$. Under these conditions, the mixed state becomes the global minima ($F(E)_{mixed} = - 52.17$), and we find a large energy barrier to the next local minima corresponding to a pure C-CDW state ($F(E)_{C} = - 5.66$). On the contrary, a much larger value of $d_0 = 6.1$ leads, Fig. \ref{fig:f4}\textbf{d-e}, into a state in which an almost pure H-CDW phase emerges as a local minimum ($F(E)_H = -2.28$), although with a much shallower energy barrier ($F(E)_C = -5.66$). We note that $d_0$ encodes the number and size of adjacent domains with different order parameters that emerges during the quench and it is proportional to the length of neighboring domain walls of the two competing orders. Thus, large values of $d_0$ are associated with larger populations of small domains, expected in fast quenches. Small values of $d_0$ indicate the existence of larger domains, consistent with intermediate quenches. We note that for values of $d_0 < 0.5$, competition is not possible, and there is only a global minimum corresponding to the C-CDW phase within our model. The dependence of the global minima on $d_0$ and $a_0$ is shown in Fig. \textbf{f}, where the color encodes $F(E)_{mixed}/F(E)_H$.

Our data is consistent with a scenario in \tas{} in which a thermal quench stabilizes a mixed phase of H-CDW and C-CDW order parameters up to $T\!=\! 180$~K in cooling and $T\!=\! 210$~K in warming. We note that coexistence of the H-CDW and C-CDW phase in ultrafast switched samples has also been recently reported \cite{burri2024}. Moreover, all our thermal quenching attempts were unsuccessful in engineering a sample without signatures of the C-CDW state indicating that sufficiently large values of $d_0$ where not achieved. It is likely that the quench protocol we describe here results in a nonuniform quench rate that is locally too slow over some regions of the sample to allow for the nucleation of the H-CDW domains. Within this context, previous ultrafast stabilization of the H-phase in thin flakes \cite{Stojchevska_Science_2014,Stahl_NatComm_XRD,Vaskivkyi_2015, hollander2015electrically,ma2016metallic,cho2016nanoscale,Venturini_2022,Tsen_pnas}, can be understood as a non-thermal ultrafast quench \cite{PhysRevX.10.021028,masoumi2024} leading to high domain densities and large values of $d_0$. Our results are consistent with proposals in which the absorption of energy by the electronic degrees of freedom suppresses the C-CDW order parameter. The consequent relaxation of the non-thermal electrons via the lattice leads to an ionic displacement from that of the C-CDW state into that of the H-CDW \cite{Retting_2023} allowing for the correlations of the H-CDW order parameter to grow \cite{PhysRevX.10.021028}.  Thus, our results and Landau-Ginzburg theory calculations suggest that the minimum of the free energy landscape associated with the H-CDW order parameter already exists at the thermodynamic equilibrium of \tas{} and can be accessed with a sufficiently fast adiabatic quench (Figure \ref{fig:f4} \textbf{a}). 

Our quenching protocol and the observation of the mixed-CDW phase opens a new pathway towards the development of new efficient functional devices based on \tas{} in which H and C-CDW domain engineering can lead to switchable metallic behavior at temperatures above LN$_2$ \cite{tas2_devices,Mraz2022,Venturini_2022}. More generally, our results have implications for the dynamic control of quantum materials \cite{RevModPhys.93.041002,jarc2023cavity} by establishing that for macroscopic order parameters, the engineering of free energy landscapes can be achieved without the need for strong laser pulses when a thermally stable phase is already present at equilibrium \cite{KZ_problem,KB_mechanism_Ising,Wu2024}.

\section{Data Availability}
The datasets obtained in this study are available in the open database at (To be provided).

\section{Acknowledgements}
A.d.l.T acknowledges helpful conversation with Michael Buchold, Yue Cao, Martin Claassen, Simon Gerber, and Dante M. Kennes. Work performed at Brown University by A.d.l.T., Q.W. and K.W.P. was supported by the U.S. Department of Energy, Office of Science, Office of Basic Energy Sciences, under Award Number DE-SC0021223. P.M.V., S.M.H., and D.B. acknowledge support from NSF DMR 2226097 and the Mason Graduate Division’s Presidential Scholarship Program. This work is based on research conducted at the Center for High-Energy X-ray Sciences (CHEXS), which is supported by the National Science Foundation (BIO, ENG and MPS Directorates) under award DMR-1829070.  G.A.F. and Y. M. gratefully acknowlege support from NSF DMR-2114825, and G.A.F. from the Alexander von Humboldt Foundation. 

\section*{Author Contributions}

A.d.l.T and Q. W. synthesized and characterized the single crystal samples.  A.d.l.T., Q. W., K. W. P., and J.P.C.R. performed the High Dynamic Range x-ray scattering measurements.  All x-ray experimental data was analyzed and interpreted by A.d.l.T and  K~.W.~P.. B. C., J. V. R., and S. M. H. performed and analyzed the STM and STS measurements. D. B. and P.M.V. performed the cryo-Raman experiments. Y. M. and G.A.F. performed the Landau-Ginzburg free energy calculations. A.d.l.T. and. K.~W.~P wrote the manuscript with input from all co-authors.

\section{Methods}\label{Methods}

\subsection{Sample Growth}

\tas{} bulk single crystals were grown by chemical vapor transport, with iodine as the transport agent. Stoichiometric amounts of elemental Ta and S were sealed in a quartz tube. The tubes were held at a $950^{\circ}$~C – $850^{\circ}$~C temperature gradient for 240 h and then placed in ice water to stabilize the 1T phase. Samples were characterized by XRD, resistivity, and Raman (see Fig. \ref{fig:E2}).

\subsection{Thermal quench}

Bulk \tas{} crystals were sealed in a 1$^{"}$ quartz tube filled with Ar gas. The \tas{} samples were then annealed by warming them up to 420~K, below the polytype transition at 660~K, with an 8h hold. This was followed by either a fast quench in which the quartz tube is dropped in a reservoir of LN$_2$ or slow cooling at room temperature. In-situ quench was performed at the QM2 beamline of the CHESS synchrotron by alternating between hot gas and a LN$_2$ cryo-stream.

\subsection{Estimate of the cooling rate}

To provide an estimate of the cooling rate for our \tas{} sample quenched within an Ar-filled quartz ampule shown in Fig. \ref{fig:f4} \textbf{a} and Extended data, we solve the heat diffusion equation for a system with cylindrical symmetry under the assumption that \tas{} is at all times thermalized with the Ar bath and that the outside surface of the quartz tube is thermalized with the bath. Thus, the limiting factor to thermalization is the heat diffusion from the hot Ar volume at the starting temperature ($\rho{Ar} = 1.8$ kg/m$^{3}$, $K_{Ar} = 0.0179$ W/(m·K), $C_p^{Ar} = 400$ Jkg$^{-1}$m) through the $1.1$~mm thick quartz wall with the outer surface acting as the heat sink ($\rho_{Q} = 2650$ kg/$m^{3}$, $K_{Q} = 1$,$C_p^{Q} = 700 $ JKg$^{-1}$m)).

\subsection{X-ray High Dynamic Range Reciprocal Space Mapping}

Hard x-ray (20 keV) HDRM was performed at the QM2 beamline of Cornell High Energy Synchrotron Source (CHESS) using a Pilatus 6M area detector. The sample is mounted on a Kapton pin using GE varnish and rotated around three different axes by 360$^{\circ}$. Data is collected at every 0.1$^{\circ}$ step with a frame rate of 0.1s. The data is reduced into a 3D stack, indexed, and projected into a 2D dataset using the beamline software. 

\subsection{In-house XRD measurements}
In-house XRD measurements were performed on a Huber four-circle diffractometer in combination with a micro-focused Ag $K_{\alpha}$ x-ray source and a 256 x 256 pixel GaAs detector (pixel size 55 x 55 $\mu$m) situated 96 cm away from the center of rotation. Data was reduced using the same software as for the CHESS data.

\subsection{STM/STS measurements}
STM and STS measurements were performed on a RHK Technology PanScan Freedom closed-cycle STM. Samples for the STM/STS measurements were mounted on stainless steel post holders using conductive silver epoxy and quenched as previously described. After the quench, the samples were characterized by XRD and kept below $T = 370$K. The surface of a \tas{} crystal was cleaved at room temperature and $\approx 10^{-10}$~mbar by carbon tape and a cleaving screw before being introduced to the STM stage and cooled to 10~K. All data presented here were taken at 10~K. 

\subsection{Resistivity}

Resistivity measurements were performed using a Quantum Design Physical Properties Measurement System (PPMS). Samples were mounted on a DC resistivity puck and a standard four-probe method was used, with Au wires and Ag paste contacts. 

\subsection{Raman Spectroscopy}
Raman measurements of quenched and unquenched TaS$_2$ single crystals are performed at 5 K in a closed-cycle He cryostat. The samples are excited in a backscattering geometry with a 532 nm laser focused through a 0.6 NA objective lens with 40$\times$ magnification. The laser power was measured to be 110 $\mu$W before the objective. The scattered light was collected through the same lens and directed to an imaging spectrograph with a liquid nitrogen-cooled charge-coupled device. Rejection of the Rayleigh scattered light is accomplished using three-volume Bragg gratings that allow measurement of Raman photons down to 10 cm$^{-1}$.

%

\newpage
\begin{widetext}
\section{Extended Data}

\setcounter{figure}{0}
\renewcommand{\thefigure}{{\bf \arabic{figure}}}
\renewcommand{\figurename}{{\bf Extended Data}}
\renewcommand{\thetable}{Supplementary \arabic{table}}
\renewcommand{\theequation}{S\arabic{equation}}
\renewcommand{\bibsection}{}
\begin{figure}[!hp]
    \includegraphics[width=0.7\textwidth]{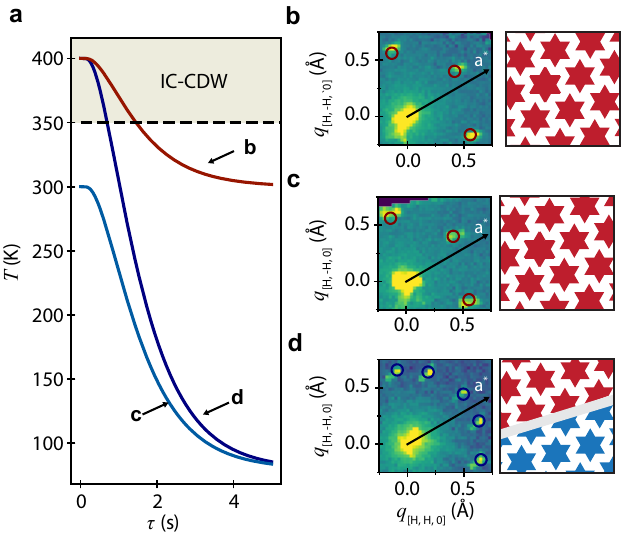}
	\caption{\textbf{a} Cooling rate simulations and reciprocal space maps and CDW state schematic \textbf{b-d} for three different scenarios: \textbf{b} slow cool from 400 K, \textbf{c} quench from 300 K and \textbf{d} quench from 400 K. Only the third case shown in \textbf{d} results in the emergence of the mixed CDW state.}
	\label{fig:E1}
\end{figure}
\begin{figure}[!hp]
    \includegraphics[width=.7\textwidth]{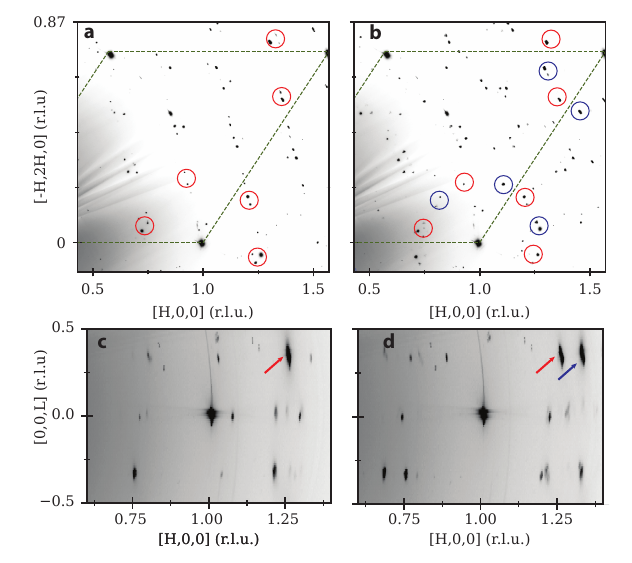}
	\caption{\textbf{a}. Reciprocal space maps parallel to the [HK0] plane ($\delta L \!=\! .5$, L \!=\! 0) at $T \!=\! 300$~K for an as-grown sample. \textbf{b}. Reciprocal space maps parallel to the [HK0] plane ($\delta L \!=\! .5$, L \!=\! 0) at $T \!=\! 300$~K after a quench. Red and blue markers highlight the original NC-CDW and the second set of satellite peaks that emerge after the quench, respectively. Green dashed lines highlight the $a^{*}$ and $b^{*}$ directions. \textbf{d} Reciprocal space map parallel to the [H0L] plane for the as-grown sample ($\delta K \!=\! .15$, K \!=\! 0). \textbf{e} Reciprocal space map parallel to the [H0L] plane for the quenched sample ($\delta K \!=\! .15$, K \!=\! 0). Red and blue arrows point to two sets of out-of-plane Bragg peaks corresponding to the NC-CDW. The L dependence and central value of these peaks are different than those seen in the mixed phase sample for the H- CDW phase.}	\label{fig:E2}
\end{figure}
\begin{figure}[!p]
    \includegraphics[width=0.7\textwidth]{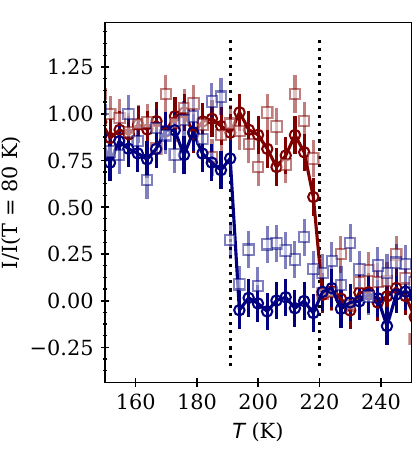}
	\caption{Circular and square markers show the temperature dependence in cooling (blue) and warming (red) of two CDW Bragg peaks of opposite chirality in quenched samples of \tas{}}
	\label{fig:E3}
\end{figure}
\begin{figure}[!p]
    \includegraphics[width=0.7\textwidth]{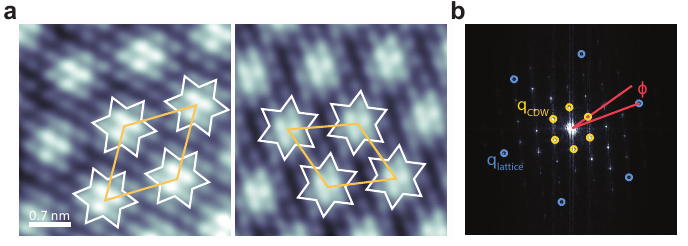}
	\caption{\textbf{a} STM data with atomic resolution in two domains with different chirality in Fig. \ref{fig:f3}\textbf{b}. \textbf{b} The high-intensity fundamental peaks ($q_{lattice}$) are highlighted in blue. The satellite peaks ($q_{CDW}$, yellow) are associated with the TH-CDW phase. The resolution of our STM measurements prevents us from resolving the difference in the magnitude and in-plane rotation of $|q|$ between the C-CDW and H-CDW.}
	\label{fig:E4}
\end{figure}
\begin{figure}[!h]
    \includegraphics[width=0.7\textwidth]{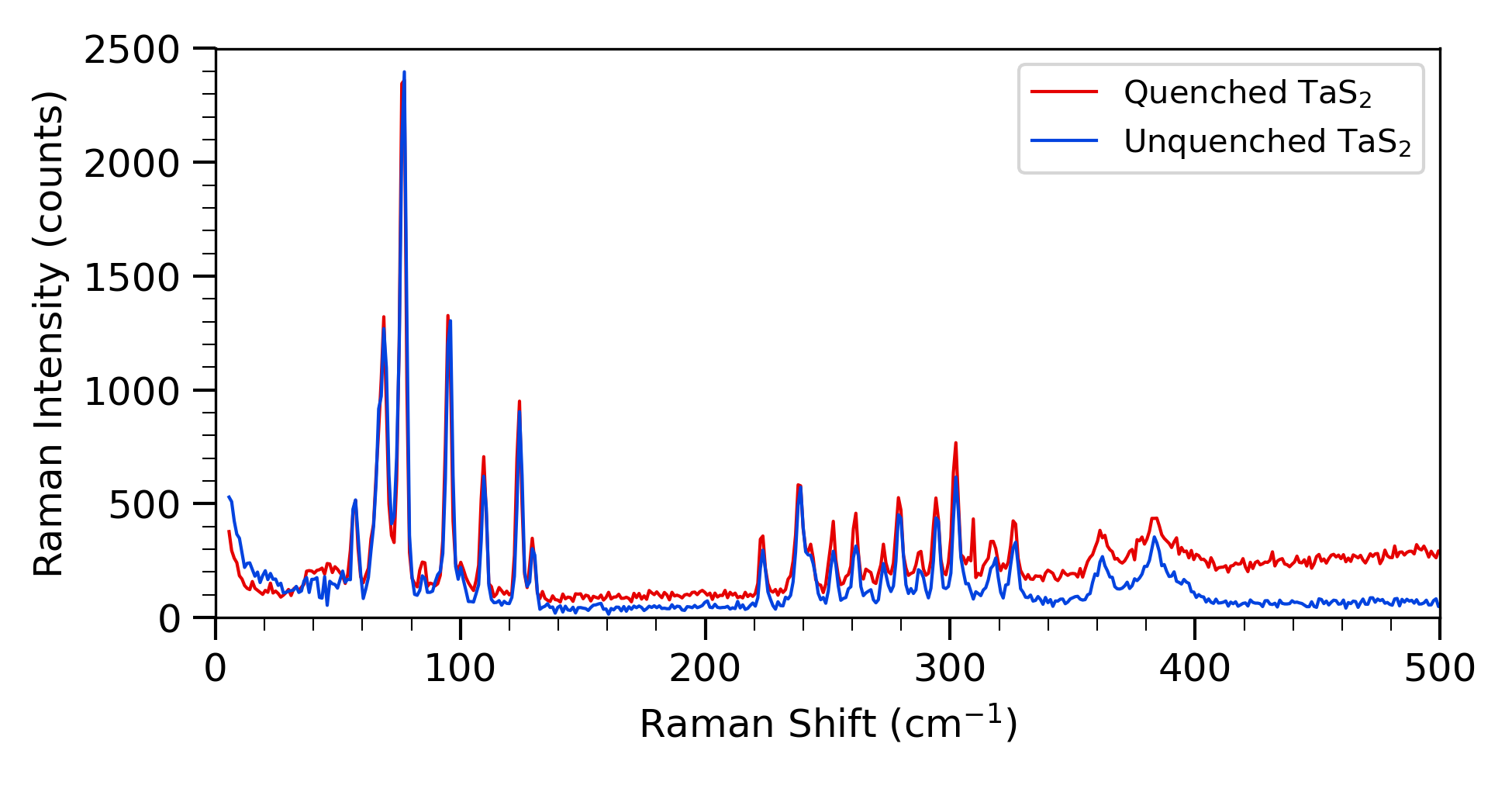}
	\caption{Unpolarized Raman spectra in a quenched (red) and as-grown (blue) at $T \!=\! 5$~K.}
	\label{fig:E5}
\end{figure}
\end{widetext}
\end{document}